\documentclass[conference]{IEEEtran}
\usepackage{cite} 
\usepackage{amsmath,amssymb,amsfonts} 
\usepackage{graphicx} 
\usepackage{textcomp} 
\usepackage{xcolor} 
\usepackage[utf8]{inputenc}
\usepackage{amsmath}
\usepackage{multirow}
\usepackage{array}

\usepackage{cite}
\usepackage{amsmath,amssymb,amsfonts}
\usepackage{graphicx}
\usepackage{textcomp}
\usepackage{graphicx} 
\usepackage{subcaption}
\def\BibTeX{{\rm B\kern-.05em{\sc i\kern-.025em b}\kern-.08em
    T\kern-.1667em\lower.7ex\hbox{E}\kern-.125emX}}

\usepackage{algorithm} 
\usepackage{algpseudocode} 

\begin{document}

\title{Non-Invasive Monitoring of Vital Signs in Calves Using Thermal Imaging Technology}

\author{\IEEEauthorblockN{Ehsan Sadeghi}
\IEEEauthorblockA{\textit{EEMCS Faculty} \\
\textit{University of Twente}\\
Enschede, The Netherlands \\
e.sadeghi@utwente.nl}
\and
\IEEEauthorblockN{Zinan Guo}
\IEEEauthorblockA{\textit{EEMCS Faculty} \\
\textit{University of Twente}\\
Enschede, The Netherlands \\
z.guo-1@student.utwente.nl}
\and
\IEEEauthorblockN{Alessandro Chiumento}
\IEEEauthorblockA{\textit{EEMCS Faculty} \\
\textit{University of Twente}\\
Enschede, The Netherlands \\
a.chiumento@utwente.nl}
\and
\IEEEauthorblockN{Paul Havinga}
\IEEEauthorblockA{\textit{EEMCS Faculty} \\
\textit{University of Twente}\\
Enschede, The Netherlands \\
p.j.m.havinga@utwente.nl}
}

\maketitle

\begin{abstract}
This study presents a non-invasive method using thermal imaging to estimate heart and respiration rates in calves, avoiding the stress from wearables. Using Kernelised Correlation Filters (KCF) for movement tracking and advanced signal processing, we targeted one ROI for respiration and four for heart rate based on their thermal correlation. Achieving Mean Absolute Percentage Errors (MAPE) of 3.08\% for respiration and 3.15\% for heart rate validates the efficacy of thermal imaging in vital signs monitoring, offering a practical, less intrusive tool for Precision Livestock Farming (PLF), improving animal welfare and management.

\end{abstract}


\section{Introduction}
\label{sec:introduction}

Cattle play a crucial role in global agriculture, pivotal in dairy and beef production and essential for food supplies. Their health and well-being are central to Precision Livestock Farming (PLF), which aims to minimize stress and align with ethical standards for better productivity. Monitoring vital signs like heart rate (HR) and respiration rate (RR) is vital for managing health and preventing disease proactively. Newly born calves, highly vulnerable with higher early mortality rates, often suffer from diseases marked by significant vital sign changes, highlighting the importance of early and effective monitoring \cite{myref-5}.

Recently, there has been a shift towards non-contact animal health monitoring, employing technologies such as camera systems, radar, and thermal imaging \cite{sadeghi2024raypet, animal-pig}. These methods, particularly thermal imaging, reduce monitoring stress by detecting subtle body temperature changes reflecting physiological processes without direct contact. Traditional wearables, which can be stressful, especially for recently born animals prone to stress and health complications, are less ideal in comparison \cite{MySurvey}. Thermal cameras track dynamic temperature fluctuations, capturing the rhythm of blood flow and respiratory cycles through thermal videos. This allows for non-invasive observation of HR and RR, making thermal imaging a crucial tool for monitoring vital signs in vulnerable newborn animals with high mortality risks.

Thermal cameras have become essential in human vital sign monitoring, surpassing standard cameras by capturing temperature variations invisible in RGB images. This technology has broad applications in human health, such as monitoring RR in cyclists and neonates \cite{human-3, human-15}. In veterinary science, its use is burgeoning. For example, Jorquera-Chavez et al. used thermal and RGB cameras to monitor HR and RR in cattle, finding a strong correlation for RR but variable accuracy for HR, underscoring challenges in consistent HR measurement \cite{animal-cattle}. Similarly, Pereira et al. demonstrated its application in pigs, although their animals were anesthetized, limiting the study's relevance to typical farm conditions \cite{animal-pig}. These cases illustrate the potential and current limitations of thermal imaging in animal health, highlighting the need for robust studies that reflect real-world settings.

Although remote monitoring of animal vital signs is not new, the focus on newborn farm animals like calves and piglets is scant. Their high mortality and the drawbacks of wearables underscore the urgent need for non-invasive monitoring from birth. This study is the first to use thermal imaging for this purpose, employing advanced imaging and signal processing to compare its accuracy against traditional methods. Our research aims to fill a crucial gap, offering insights that could significantly advance non-invasive techniques in PLF.

\begin{figure*}
\centering
\includegraphics[width=18 cm]{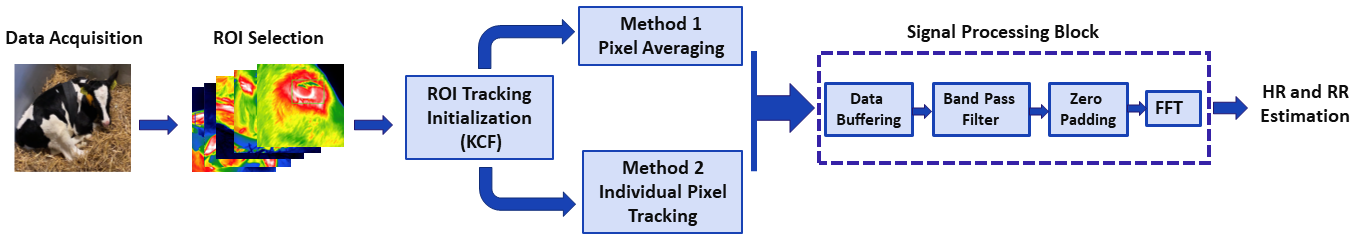}     
\caption{Block diagram illustrating the process flow for estimating heart rate and respiration rate from thermal images of calves.}
\label{BD}
\vspace{-0.2cm}
\end{figure*}

\section{Methodological Framework}


Our methodology for thermal imaging-based vital sign estimation is streamlined into a few key steps, as depicted in Figure \ref{BD}. The process begins with thermal data acquisition from the calves, followed by the selection of specific Regions of Interest (ROIs). We utilize the Kernelized Correlation Filters (KCF) for robust ROI tracking. Data is then processed using two methods: one averages pixel values, and the other tracks individual pixels for detailed analysis. Both methods employ signal processing techniques like filtering and Fourier Transform to determine HR and RR, efficiently summarizing our comprehensive approach.

\subsection{Video/ Image Acquisition}

Our study utilizes the Seek Thermal Compact Pro camera, which records thermal videos via a smartphone connection. This technology captures infrared radiation, reflecting temperature variations essential for monitoring physiological changes. The camera's high resolution and sensitivity enable it to detect subtle thermal fluctuations—like those from calf inhalation and exhalation or cardiac activity—that signify vital signs such as HR and RR. We focus on changes over time rather than absolute temperatures, analyzing pixel intensity shifts to accurately estimate HR and RR. The chosen frame rate ensures precise capture of these rapid physiological changes, crucial for reliable data analysis.

\subsection{Region of Interest}

\begin{figure}[t]
\centering
\includegraphics[width=8.9 cm]{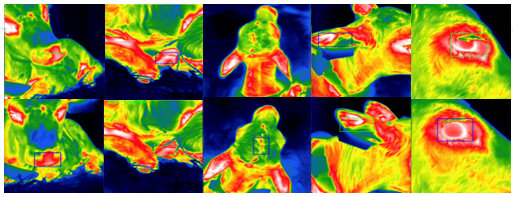}     
\caption{ROIs on a calf identified for vital signs monitoring: nose, forehead, hoof, eye, and ear areas.}
\label{ROI}
\vspace{-0.3cm}
\end{figure}

Effective thermal imaging for physiological monitoring relies on precisely identifying and tracking Regions of Interest (ROIs) that show significant thermal fluctuations due to biological processes. In this study, ROIs were strategically selected on the calves' bodies to enhance signal clarity and provide meaningful physiological insights.

For respiration rates, the nose area was chosen as the ROI because it distinctly displays temperature changes with breathing cycles, captured from multiple angles—frontal and lateral—to analyze the respiration cycle thoroughly. For heart rate, the forehead was selected based on its correlation with internal physiological states, where superficial blood vessels show visible thermal shifts with each heartbeat \cite{ROI-1}. Additional areas like the hooves, eyes, and ears were also monitored for their thermal variations during cardiac activity, aiding in a comprehensive assessment.

Thermal videos of these regions were recorded throughout the study, allowing for a detailed comparative analysis of their effectiveness in estimating vital signs. Figure \ref{ROI} illustrates these ROIs, underscoring the targeted areas for optimal measurement accuracy and reliability.

\subsection{Tracking the ROIs}
Consistent tracking of the designated ROIs is critical throughout the thermal video sequences to accommodate the calves' movement. Initially, ROIs are manually identified in the first frame, but continuous manual re-identification is impractical for automated analysis. Thus, robust tracking algorithms are essential for maintaining accurate and reliable vital sign monitoring.

We evaluated the Kanade-Lucas-Tomasi (KLT) and Kernelized Correlation Filters (KCF) tracking algorithms \cite{KCF}. KLT, while effective in high-texture environments, struggles with the minimal texture and pronounced noise in thermal imagery, often requiring manual corrections. In contrast, KCF utilizes a kernel trick to enhance tracking performance in varied conditions, making it better suited for the subtle and homogeneous thermal patterns in our study.

After testing both, KCF's superior ability to maintain consistent tracking without the need for high-contrast features led to its selection as the primary tracking algorithm. It proved highly effective, reliably tracking the ROIs through smoothly varying thermal patterns, and is thus the chosen method for ensuring the accuracy of our physiological measurements.


\subsection{Signal Pre-processing and Vital Sign Estimation}

Our thermal camera provides us with a frame rate of 15 frames per second (FPS), equivalent to a 15 Hz sampling frequency. Considering the normal range of heart rate and respiration rate for calves, which is 100 to 160 beats per minute (BPM) or 1.67 to 2.67 Hz, and 10 to 40 respiration per minute (RPM) or 0.17 to 0.67 Hz, Nyquist’s theorem suggests that the sampling frequency should be at least $2\times 2.67=5.34$ Hz. This requirement is comfortably met by our thermal camera’s sampling rate, allowing for successful estimation of both heart rate and respiration rate.

To estimate heart rate and respiration rate from the thermal videos, we employ two distinct methods for processing the pixel values (intensities) within the ROIs. Both methods involve initial data acquisition from the ROIs, followed by digital signal processing techniques to filter and analyze the frequency components.

\subsubsection{Method 1. Average Pixel Value Analysis}

In the first method, we simplify the data by calculating the average pixel intensity within the ROI for each frame. This approach reduces the data set to a single time series, representing the average thermal signal over time across the entire ROI. The average pixel values for each frame are stored in a buffer—a temporary holding place that allows for continuous data input and sequential processing. This buffer is then subjected to a series of signal-processing steps.
A band-pass filter (BPF) is applied to isolate the frequency bands pertinent to HR and RR. This filter removes high-frequency noise and very low-frequency trends that are irrelevant to the physiological signals of interest.
To improve the resolution of the frequency analysis, zero padding is applied to the buffered signal before the Fourier transformation. This step involves extending the signal with zeros, thereby increasing the total length of the data without altering its statistical properties.
The extended signal undergoes FFT to convert it from the time domain to the frequency domain. The dominant frequency within the predefined HR and RR frequency bands is identified as the peak frequency, representing the estimated physiological rate.

\subsubsection{Method 2. Individual Pixel Tracking Analysis}
The second method tracks each pixel value within the ROI across all frames. This method retains a more comprehensive dataset, capturing individual pixel fluctuations over time, which could contain more detailed information about the physiological processes. Similar to Method 1, these pixel values are also stored in a buffer; however, here, each pixel's time series is processed independently.

Each pixel series is filtered using the same BPF criteria to focus on the HR and RR frequency ranges.
Each filtered pixel series is zero-padded and transformed using FFT.
Instead of a single dominant frequency, this method examines the frequency spectrum of each pixel and selects the most commonly occurring dominant frequency across all pixels within the HR and RR ranges. 

By utilizing these two approaches, we aim to compare the effectiveness of aggregate versus detailed pixel analysis in extracting reliable physiological signals from thermal imagery. This comparison will help determine which method provides a more accurate reflection of the vital signs in calves, contributing to better health monitoring practices.

\section{Experimental Design}

To validate our methodology, we gathered data from four calves aged between 6 to 10 weeks on a farm near Eibergen, the Netherlands. The calves were housed in individual pens, a common practice in the early weeks post-birth, to minimize stress and potential interference.

We used a Seek Thermal Compact Pro camera for recording thermal images. This user-friendly device, which pairs with a smartphone, offers a cost-effective solution for thermal imaging, making it an attractive option for farmers. The camera boasts a $320 \times 240$ resolution and a thermal sensitivity of less than 70 mK. With its 32-degree field of view, the camera facilitates non-invasive recording from a distance of approximately 1 meter, avoiding unnecessary stress to the animals. The recording durations were standardized at 60 seconds for capturing the respiration and heart rate ROIs.

\begin{figure}
\centering
\includegraphics[width=8 cm]{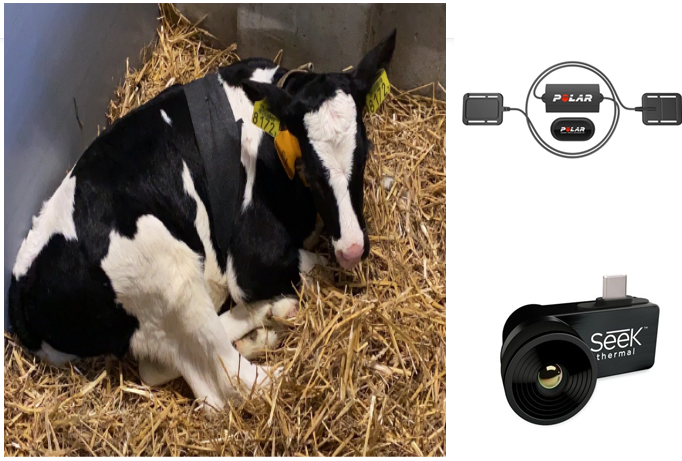}     
\caption{A calf with a heart rate sensor, the Seek Thermal camera, and the Polar Electrode Set used in the study.}
\label{setup}
\vspace{-0.3cm}
\end{figure}

For ground truth validation of heart rate measurements, we employed the Polar Equine H3 Heart Rate Sensor Electrode Set, which has been tested and validated for bovine use in previous studies \cite{polarthing}. This particular model offers adjustable electrode distances, enhancing the robustness and accuracy of heart rate detection. The electrodes were positioned and then secured with straps to ensure close skin contact. To enhance conductivity, the calf's skin was shaved, and a conductive gel was applied before electrode placement.
For verifying the respiration rate, we utilized a high-resolution camera aimed at the calf's abdominal area to visually count the inhalation and exhalation cycles, allowing for an accurate measurement of breathing frequency.
Figure \ref{setup} illustrates the image of the experimental setup featuring three key components: a calf in an individual pen wearing the Heart Rate Sensor, the Seek Thermal camera used for recording, and the Polar Equine H3 Heart Rate Sensor Electrode Set.

\section{Results}

For each ROI, a cumulative total of 15 minutes of data were recorded across all four calves. Following this, We randomly selected 20 and 30-second segments from each minute for signal pre-processing to estimate heart and respiration rates, respectively.
The aggregated results for each ROI, representing data from all four calves, are listed in Table \ref{results}, where the performance of both Method 1 and Method 2 is contrasted.

\begin{table}[h!]
\centering
\begin{tabular}{|c|c|c|c|c|}
\hline
\multirow{2}{*}{Vital Signs} &\multirow{2}{*}{ROIs} & Num. of  & \multicolumn{2}{c|}{MAPE} \\
\cline{4-5}
&& Recordings& Method 1 & Method 2 \\
\hline
Heart Rate & Forehead & 15 & 10.78\% & 3.15\% \\
\hline
Heart Rate & Ear & 15 & 7.43\% & 10.82\% \\
\hline
Heart Rate & Eye & 15 & 11.23\% & 5.23\% \\
\hline
Heart Rate & Hoof & 15 & 10.47\% & 5.63\% \\
\hline
Respiration Rate & Nose & 15 & 3.73\% & 3.08\% \\
\hline
\end{tabular}
\caption{Comparative Analysis of Method 1 and Method 2 for Estimating Heart and Respiration Rates Across Different ROIs in Calves.}
\label{results}
\end{table}
\begin{figure*}[h]  
    \centering
    \begin{subfigure}[b]{0.49\textwidth}  
        \includegraphics[width=\textwidth]{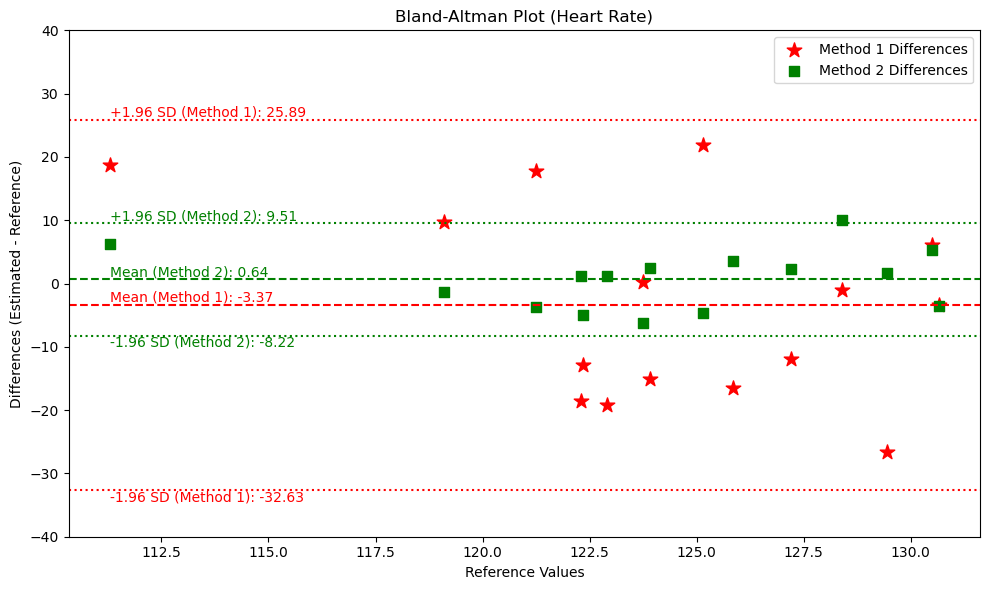}
        \caption{Heart Rate Estimation }
        \label{bland-hr}
    \end{subfigure}
    \hfill 
    \begin{subfigure}[b]{0.49\textwidth}
        \includegraphics[width=\textwidth]{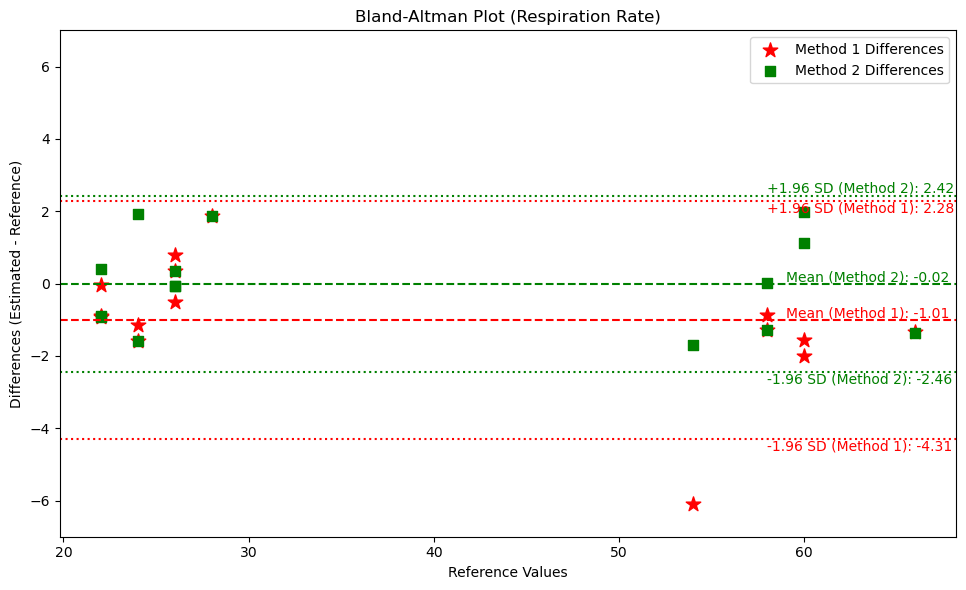}
        \caption{Respiration Rate Estimation }
        \label{bland-rr}
    \end{subfigure}
    \caption{Bland-Altman plots for HR and RR estimation methods, showcasing variance and mean differences in measurement accuracy.}
    \label{bland}
\end{figure*}

The table reveals that the nose area ROI provides the lowest MAPE values, with 3.08\% for Method 1 and 3.73\% for Method 2, indicating high accuracy in respiratory rate estimation. In the assessment of heart rate, the forehead ROI shows the most promise, exhibiting a MAPE of 3.15\% for Method 2. Taking into account the entire dataset, Method 2, which analyzes all pixel data, outperforms Method 1, which relies on the evaluation of average pixel intensity.

To validate these observations, the levels of agreement for Mehod 1 and Method 2 estimations, particularly for the nose area and forehead ROI, were subjected to Bland-Altman analysis, as illustrated in Figure \ref{bland}.

As for heart rate estimation using the forehead ROI (Fig. \ref{bland-hr}), Method 1 displays a greater variance with a broader spread of data points between the limits of agreement (LoA), as evidenced by the wider gap between the +1.96 SD and -1.96 standard deviation (SD) lines. The plot suggests a larger systematic bias and less consistency compared to Method 2, which demonstrates a tighter clustering of data points and narrower LoA, indicating more consistent estimations in line with reference values.
In the case of respiration rate estimation from the nose area ROI (Fig. \ref{bland-rr}), Method 1 presents a larger mean difference, suggesting less agreement with reference values on average compared to Method 2. Moreover, the larger spread of differences implies less precision. Method 2, with a slight mean difference hinting at a minor systematic bias, shows a more compact grouping of differences, suggesting a more precise technique, even if it slightly over- or underestimates the rates.

In both plots, the presence of outliers is noticeable, particularly for Method 1 in the heart rate plot, which could be indicative of instances where this method is less reliable. It's also worth noting that the scale of differences in the heart rate plot is considerably larger than in the respiration rate plot, suggesting that heart rate measurements may be more prone to variation between the two methods compared to respiration rate measurements.
Overall, Method 2 seems to offer a more consistent and precise estimation for both heart and respiration rates, as evidenced by the denser clustering of data points within the LoA in the Bland-Altman plots. The choice between methods will ultimately depend on the level of accuracy and precision required to ensure calf wellbeing or for the intended clinical application.


\section{Discussion}

The study demonstrates that the Seek Thermal camera can be successfully utilized for animal welfare assessment in calves, offering a non-invasive, reliable, and stress-free approach to vital sign monitoring. We investigated various ROIs to identify the most effective area for heart rate estimation. While the forehead region yielded the most promising results, outcomes may vary with the use of different, possibly more advanced thermal cameras. Nevertheless, more sophisticated thermal cameras are often prohibitively expensive, presenting an impractical solution for farmers. This reality underpins our decision to employ an affordable, portable thermal camera as a pragmatic option for PLF.
For the broader application of this technology, a more comprehensive assessment involving a larger number of calves is necessary. The restricted availability of recently born calves on the farm posed a limitation in this study, resulting in a smaller dataset than desired.

Evaluation of the Mean Absolute Percentage Error (MAPE) across all ROIs and both methods indicates that Method 2, which considers all pixel values, is more accurate, displaying low error rates and minor systematic bias without evidence of proportional bias. It is important to note, however, that the success of Method 2 heavily relies on the performance of the tracking algorithm to ensure consistent analysis of approximately the same pixels over time. Therefore, the selection of a robust tracker is critical for the efficacy of this proposed method in PLF scenarios.

\section{Conclusion}

The study demonstrates thermal imaging as a viable, non-invasive tool for monitoring calves' vital signs, providing a stress-free alternative to traditional methods. Our findings reveal that thermal imaging not only provides high accuracy in vital sign measurement, as evidenced by low MAPE values, but also introduces a method that can be seamlessly integrated into current livestock management practices to improve animal welfare and operational efficiency. While our results are promising, they also highlight the necessity for further research with a larger sample size to generalize the findings and enhance the robustness of the technology. Ultimately, this approach holds the potential to transform PLF by providing a reliable, cost-effective, and ethically preferable method for health monitoring in farm animals. Future studies should aim to refine the technology and expand its application to include a wider range of animal species and farm conditions.

\bibliographystyle{IEEEtran}
\bibliography{main.bib} 
\end{document}